\documentstyle[psfig,onecolumn]{mn2e}
\newif\ifAMStwofonts

\newcommand{\etal}{{et al.} }

\newcommand{\mcarlo}{{Monte Carlo} }

\newcommand{\lossym}{{$D[(a/c),N_{H},E]$} }
\newcommand{\lossymp}{{$D[(a/c),N_{H},E]$}}
\newcommand{\losstotf}{{$CL[(a/c),N_{H},\Gamma,E_{U}]$} }
\newcommand{\losstotfp}{{$CL[(a/c),N_{H},\Gamma,E_{U}]$}}
\newcommand{\proxyratio}{{$PX[(a/c),N_{H},\Gamma]$} }
\newcommand{\proxyratiop}{{$PX[(a/c),N_{H},\Gamma]$}}

\newcommand{\chandra}{{\it Chandra} }

\newcommand{\fekalfa}{{Fe~K$\alpha$} }
\newcommand{\fekalfap}{{Fe~K$\alpha$}}

\newcommand{\fekbetap}{{Fe~K$\beta$}}
\newcommand{\nika}{{Ni~K$\alpha$} }

\newcommand{\nh}{$N_{\rm H}$ }
\newcommand{\nhp}{$N_{\rm H}$}

\newcommand{\thetaobs}{{$\theta_{\rm obs}$} }
\newcommand{\thetaobsp}{{$\theta_{\rm obs}$}}

\newcommand{\tablecosrange}{{Table~1} }

\newcommand{\figelossfunc}{{Fig.~1} }
\newcommand{\figelossfuncp}{{Fig.~1}}
\newcommand{\figelossac}{{Fig.~2} }
\newcommand{\figelossacp}{{Fig.~2}}

\newcommand{\figelosscum}{{Fig.~3} }
\newcommand{\figelosscump}{{Fig.~3}}
\newcommand{\figtotvsnh}{{Fig.~4} }

\newcommand{\figtotvsac}{{Fig.~5} }

\newcommand{\figtotvsgam}{{Fig.~6} }

\newcommand{\figproxyvsnh}{{Fig.~7} }

\newcommand{\figproxyvsnha}{{Fig.~7(a)} }

\newcommand{\figproxyvsnhb}{{Fig.~7(b)} }

\title{The energy budget for X-ray to infrared reprocessing in
Compton-thin and Compton-thick active galaxies}

\author[Tahir Yaqoob \& Kendrah D. Murphy]
{Tahir Yaqoob$^{1}$ and Kendrah D. Murphy$^{2}$ \\
$^{1}$Department of Physics and Astronomy, Johns Hopkins University, Baltimore, MD 21218. \\
$^{2}$Department of Physics, Skidmore College, 815 North Broadway, 
Saratoga Springs, NY 12866. \\
}

\date{Accepted 2010 October 28. Received Received 2010 October 27; 
in original form 2010 September 21.}

\begin{document}

\maketitle

\begin{abstract} 

Heavily obscured active galactic nuclei (AGNs)
play an important role
in contributing to the cosmic X-ray background (CXRB).
However, the AGNs found in deep X-ray surveys are often
too weak to allow direct measurement of the column density of
obscuring matter. One method adopted in recent years to
identify heavily obscured, Compton-thick AGNs under such
circumstances is to use the observed mid-infrared to X-ray luminosity
ratio as a proxy for the column density. This is based on
the supposition that the amount of energy lost by
the illuminating X-ray continuum to the
obscuring matter and reprocessed into infrared emission is
directly related to the column density and that the proxy
is not sensitive to other physical parameters of the
system (aside from contamination by 
dust emission from, for example, star-forming regions).
Using Monte Carlo simulations, we find that the energy losses
experienced by the illuminating X-ray continuum
in the obscuring matter are far more sensitive to the
shape of the X-ray continuum and to the covering factor
of the X-ray reprocessor than they are to the column density
of the material. Specifically we find that it is possible
for the infrared to X-ray luminosity
ratio for a Compton-thin source to be just as large as that
for a Compton-thick source
{\it even without any contamination from dust}. Since the intrinsic X-ray
continuum and covering factor of the reprocessor
are poorly constrained from deep X-ray survey data,
we conclude that the
mid-infrared to X-ray luminosity ratio is not a reliable
proxy for the column density of obscuring matter in AGNs
even when there is no other contribution to the mid-infrared
luminosity aside from X-ray reprocessing.
This conclusion is independent of the geometry of the obscuring matter.

\end{abstract}


\section{Introduction}
\label{irintro}

Although it has been known since the 1980s that a significant
fraction of the active galactic nuclei (AGNs) that contribute
to the CXRB X-ray background (CXRB) must be heavily absorbed
(Setti \& Woltjer 1989),
the details of the absorbed population remain uncertain
(e.g., Comastri \etal 1995; Tozzi \etal 2006; 
Polletta \etal 2006; Gilli, Comastri, \& Hasinger 2007;
Frontera \etal 2007; Mart{\'{\i}}nez-Sansigre \etal 2007;
Ballantyne \& Papovich 2009; 
Treister, Urry, \& Virani 2009;
Draper \& Ballantyne 2009;
Fiore \etal 2009;
Georgantopoulos \etal 2008, 2009; Malizia \etal 2009, 2010). 
One of the reasons for this is that the problem is highly degenerate.
Another reason is that, although much progress has been made in
the last decade or so in resolving the hard X-ray background below 10 keV,
the bulk of the resolved AGNs are too weak to 
robustly measure their column densities with X-ray spectroscopy.
As a result, a number of indirect ``proxies'' have come into use
for interpreting deep survey data and identifying heavily
obscured AGNs.
One of these is the use of the equivalent width (EW) of the
\fekalfa line at $\sim 6.4$~keV as a rough indicator of the
column density because EWs larger than $\sim 1$~keV cannot
be produced by a Compton-thin X-ray reprocessor 
(e.g., see Ghisellini, Haardt, \& Matt 1994;
Ikeda, Awaki, \& Terashima 2009; Murphy \& Yaqoob
2009, hereafter MY09). However, AGNs found in deep surveys are seldom
bright enough to allow a direct robust measurement of the
\fekalfa line EW. An additional
constraint could in principle be obtained from
an independent indicator of the intrinsic AGN continuum luminosity
such as an emission line in a different waveband.
A few have been explored, the ratio of the
luminosity of [O~{\sc iii}]~$\lambda$~5007$\AA$ luminosity to the observed X-ray
luminosity being the oldest
(e.g., Maiolino \etal 1998; Bassani \etal 1999; Zakamska \etal 2003;
Cappi \etal 2006; Panessa \etal 2006; 
Reyes \etal 2008; Bongiorno \etal 2010). 
Mel\'{e}ndez \etal (2008) have explored the use of the 
ratio of the luminosity of the [O~{\sc vi}]~$\lambda$~25.89$\mu$m emission line to
the observed X-ray luminosity as a diagnostic of the column density.
Gilli \etal (2010) have recently suggested the analogous use  
of the [Ne~{\sc v}]~$\lambda$~3426$\AA$ emission line.
However, derivation of the column density 
from these indicators is highly model-dependent as
there are degenerate solutions
(in addition to uncertainties in calibration of the relations), 
and even then, the method can only
give a line-of-sight column density, which may be different to
that out of the line-of-sight. 
A third indicator that is used to identify heavily obscured AGNs in
deep X-ray surveys is the ratio of the mid-infrared luminosity to the
{\it observed} X-ray luminosity
(e.g., Mart\'{i}nez-Sansigre \etal 2005;
Daddi \etal 2007; Hickox \etal 2007; Alexander \etal 2008;
Fiore \etal 2008, 2009; Georgantopoulos \etal 2008, 2009, 2010;
Vignali \etal 2010; Bauer \etal 2010;
Georgakakis \etal 2010, and references therein). The idea is that 
the intrinsic AGN continuum
that intercepts the obscuring matter loses energy due to
the absorption and Compton-scattering of X-ray photons and that 
this energy loss ultimately heats the medium so that it is then
reradiated in the infrared band. It is then supposed that the ratio
of the infrared to observed X-ray luminosity in AGNs 
could be an indicator
of the column density of the reprocessor. Since reprocessed
emission from dust can also provide a significant contribution to the
infrared band (e.g., Maiolino \etal 2007;
Horst \etal 2008; Rowan-Robinson, Valtchanov, \& Nandra 2009;
Georgakakis \etal 2010),
it has certainly been recognized that interpretation of the mid-infrared
to X-ray luminosity ratio indicator is complex. 
Nevertheless, it is still believed 
(e.g., Mullaney \etal 2010, and references therein) that it 
will eventually be possible to discriminate between 
different scenarios and utilize the mid-infrared to X-ray luminosity
ratio to identify Compton-thick AGNs.
In the meantime,
whilst an unusually large mid-infrared to X-ray 
luminosity ratio does not necessarily mean
that an AGN is heavily obscured, an {\it unusually small} ratio could
(it is supposed) be used to establish that an AGN is {\it not}
a candidate to be a Compton-thick source. In other
words, rejection of Compton-thick
candidates may also be useful.

Unfortunately it has never been demonstrated, even without the
complications arising from dust emission, 
that the mid-infrared emission to X-ray luminosity
ratio should be sensitive to the
X-ray absorbing column density to the extent that the
ratio could used to infer
the magnitude of that column density.
One cannot correct for the effects of dust if
the relation between the X-ray energy reprocessed
into the infrared band has not been quantified theoretically.
It is the aim of the present paper
to investigate and quantify the relationship between the energy
lost by the intrinsic X-ray continuum in the obscuring matter and the column
density of that reprocessor, {\it explicitly without
the complications of dust emission}.
Calculating the energy available from X-ray losses for reemission
in the mid-infrared band is of course only the first step in the
process. However, establishing that the energy budget from X-ray
losses is sensitive to the column density,
and {\it not} sensitive to other key parameters, is a {\it minimal}
requirement for the mid-infrared to X-ray luminosity ratio to be
a useful indicator of the column density. If we find, as we do,
that this minimal requirement is not met, then additional physical
processes can only {\it lessen} the usefulness of the diagnostic.
Our investigation is based upon the
toroidal X-ray reprocessor model of MY09, but our principal conclusions
will have general applicability and will be independent of geometry.

The paper is organized as follows.
In \S\ref{mytorusmodel}
we give a brief overview of the assumptions of the MY09 model
and describe the method we used to analyze the energy budget.
In \S\ref{elosspec} we present the results of X-ray energy losses
in the reprocessor as a function of injection energy, and
in \S\ref{cumloss} we present the results of integrating these
``loss'' spectra over an incident power-law continuum. It is
in \S\ref{cumloss} that we show the relationship between the
integrated energy loss and the column density of the
obscuring matter, as well as the dependence
of the energy losses on the covering factor and on the 
power-law spectral index. In \S\ref{irxproxy} we define
a proxy for the ratio of the infrared luminosity due to
X-ray energy losses, to observed X-ray luminosity and present the
results of computations of this ratio as a function of 
the column density of obscuring matter.
We summarize our conclusions in \S\ref{summary}.

\section{Method}
\label{mytorusmodel}

Here we give a brief overview of the critical
assumptions that our model Monte Carlo simulations are  based upon.
Full details can be found in MY09.
Our geometry is an azimuthally-symmetric
doughnut-like torus with a circular cross-section,
characterized by only two parameters, namely
the half-opening angle, $\theta_{0}$, and the
equatorial column density, \nh (see Fig.~1 in MY09).
We assume that the X-ray source
is located at the center of the torus and emits isotropically and that
the reprocessing material is uniform and
essentially neutral (cold).
For illumination by an X-ray source that is emitting isotropically,
the mean column density, integrated over all incident
angles of rays through
the torus, is $\bar{N}_{H}$~$=(\pi/4)$\nhp.
The inclination angle between the observer's line of sight and the
symmetry axis of the torus is given by
\thetaobsp, where \thetaobs$=0^{\circ}$ corresponds to a face-on
observing angle and \thetaobs$=90^{\circ}$
corresponds to an edge-on observing angle. 
In our calculations we distribute the emergent photons in
10 angle bins between $0^{\circ}$ and $90^{\circ}$ that have
equal widths in $\cos{\theta_{\rm obs}}$, and refer
to the face-on bin as \#1, and the edge-on bin as \#10
(see \tablecosrange in MY09).

We used a version of our Monte Carlo code that injects
single photons (as opposed to bundles of photons, with weights),
so that energy is explicitly conserved in the code.
Photons were injected with
energies in the range 
5--500~keV. In the present study we are interested in
the energy losses and not the emergent spectra so for
energies below 5 keV we calculated energy losses for
absorption only, since the relative contribution from 
Compton scattering is negligible in this regime.
If the solid angle subtended by the torus at the
X-ray source is $\Delta\Omega$, the covering factor
is $[\Delta\Omega/(4\pi)]=1-\cos{\theta_{\rm obs}}$ 
(\thetaobs only varies between $0^{\circ}$ and $90^{\circ}$).
The covering factor may also be
expressed in terms of the physical 
dimensions of the torus. If $a$ is the
radius of the circular cross-section of the torus, and $c+a$ is
the equatorial radius of the torus then $[\Delta\Omega/(4\pi)]=(a/c)$
(see MY09).
Our model employs a
full relativistic treatment of Compton scattering,
using the full differential and total Klein-Nishina Compton-scattering
cross-sections.
We utilized photoelectric absorption cross-sections for 30 elements as
described in Verner \& Yakovlev (1995) and
Verner \etal (1996) and
we used Anders and Grevesse (1989) elemental cosmic abundances in
our calculations.
The \fekalfap, \fekbetap, and \nika fluorescent emission lines 
were included in the model, as described in MY09.
We performed Monte Carlo simulations for $(a/c)$ in the range 0.1--1.0
and \nh in the range $3 \times 10^{22} \ \rm cm^{-2}$ to 
$10^{25} \ \rm cm^{-2}$.

\section{Energy loss spectra}
\label{elosspec}

For photons injected into the torus with an energy $E$, in the
interval $dE$, we
define $D[(a/c),N_{H},E]dE$ as the fractional energy lost by those photons
due
to X-ray absorption and Compton downscattering in the torus.
This is shown symbolically in equation~\ref{eq:elossdef},
where $F_{\rm absorbed}(E)$ and $F_{\rm C}(E)$ are fractions
of the original energy $E$ that are ultimately absorbed or
lost to the medium through Compton downscattering respectively.
$N_{0}$ is the number of photons injected but obviously cancels
out of the definition. The function $D$ does not of course
only depend on $(a/c)$, $N_{H}$, and $E$, it depends on many
other factors such as element abundances and geometry. 
A photon that undergoes Compton scattering
is ultimately either absorbed or escapes the medium. However,
absorption can result in line emission and
such line photons either escape
($F_{L}(E)$ in equation~\ref{eq:elossdef})
or are themselves absorbed. 

\begin{eqnarray}
\label{eq:elossdef}
D[(a/c),N_{H},E] & \equiv & \frac{N_{0}E[F_{\rm absorbed}(E) +F_{\rm C}(E)-F_{L}(E)]}{N_{0}E} \\ \nonumber
& = & \int_{0}^{E}{ [1-F_{\rm escape}(E,E')] \ dE'}
\end{eqnarray}

In practice, using our Monte Carlo code
we can track photons injected in a given energy interval
and count the energy that escapes. The energy that does not
escape must have been given up to the reprocessing medium
and this is represented in the 
second part of equation~\ref{eq:elossdef}. Whereas the 
functions $F_{\rm absorbed}(E)$, $F_{\rm C}(E)$, and $F_{L}(E)$ are
already integrated over the energy distribution resulting from
a single injection energy, $E$, the function $F_{\rm escape}(E,E')$
in equation~\ref{eq:elossdef} needs to be
explicitly integrated over the distribution of downscattered
energies, from $0$ to $E$.
We also summed over all escaping
angles since we are interested in the total energy lost to the
medium. Our model includes only three emission lines so we miss
some of the escaping energy and therefore overestimate the energy
loss, which is already just an upper limit on the total energy available
for reprocessing into infrared emission. 
Moreover, our conclusions in the present paper will not
be affected by neglect of the energy in the omitted emission lines.
Note that if the torus were completely opaque and non-reflective at
all energies, $D[(a/c),N_{H},E]$ would simply be equal to $(a/c)$,
the covering factor, since that is the fraction of the energy that
would be intercepted and lost to the torus.

In \figelossfunc we show curves of the quantity \lossym/(a/c) for
five different column densities in range $3 \times 10^{22} \ \rm cm^{-2}$
to $10^{25} \ \rm cm^{-2}$ (solid, black curves), computed from
our \mcarlo simulations. The curves have been
calculated for $(a/c)=0.5$ but have then been normalized by the
covering factor, $(a/c)$. The purpose
of this is to later facilitate direct comparison of the {\it shapes}
of the curves 
for different values of $(a/c)$. The units for the curves are
fractional energy loss per unit energy, per unit covering factor.
We see that at low energies, below $\sim 1$~keV, the 
fractional energy loss per keV has already reached its limiting
value of $(a/c)$ for all of the column densities shown. At higher
injection energies less of the energy is lost but the
threshold for it to drop below $(a/c)$ depends on \nhp. For the
two highest values of \nh shown ($5 \times 10^{24} \ \rm cm^{-2}$
and $10^{25} \ \rm cm^{-2}$), this energy loss drops below the
maximum only above $\sim 10$~keV.
The drop in the curves in \figelossfunc
due to the decrease in absorption with 
increasing photon injection energy
for each \nh is followed by a flattening
again and then a rise as the energy losses due to Compton downscattering
become important. For the two lowest column densities in 
\figelossfunc we have illustrated how Compton downscattering
losses compare with absorption by overlaying the equivalent
curves for pure absorption only (red curves).

We found that the {\it normalized} functions $D[(a/c),N_{H},E]/(a/c)$
could not be distinguished in
magnitude or shape (within the statistical errors of
the \mcarlo results) for any value $(a/c)$ in the range 0.1--1.0
for column densities up to $10^{24} \rm \ cm^{-2}$. For 
higher column densities a dependence of the magnitude and
shape of the fractional energy loss curves on the covering factor
does become apparent. This is shown in \figelossacp, which shows
the curves in \figelossfunc for $N_{H}=5 \times 10^{24} \rm \ cm^{-2}$
and $10^{25} \rm \ cm^{-2}$, zoomed in on the region in which
Compton scattering dominates. The black curves show the previous
results with $(a/c)=0.5$ and the red and brown curves correspond
to $(a/c)=0.1$ and $(a/c)=1.0$ respectively. The dependence
on covering factor in the Compton-thick regime arises because
the amount of energy reflected from the inner surface of the
torus and intercepted by the torus again depends on the
opening angle of the torus and therefore on the covering factor.

\begin{figure}
\centerline{
\psfig{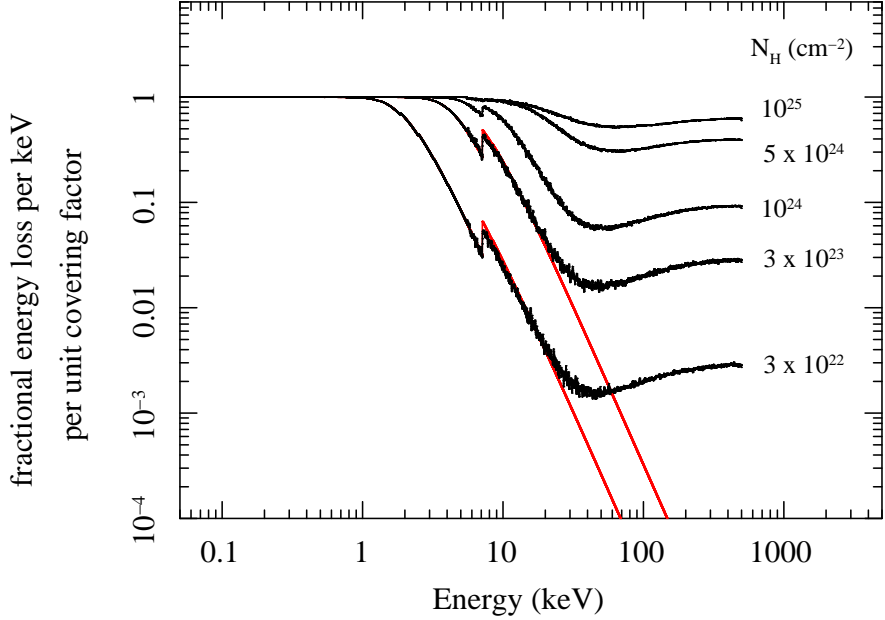}}
\caption[Fractional Energy Loss Function]{\footnotesize Curves of 
\lossym (see equation~\ref{eq:elossdef}), showing the difference between
the energy escaping the torus 
and incident energy (per keV), as a fraction of the
incident energy (per keV). The horizontal axis
corresponds to the incident energy. The curves were calculated for
a covering factor of $[\Delta\Omega/(4\pi)]=(a/c)=0.5$ but 
they are normalized by the
covering factor so that they can be directly compared with 
curves with different covering factors (e.g., see \figelossacp).
The black curves show the Monte Carlo results for five different
column densities, $N_{H}$, as shown. 
The red curves show, for
comparison, the effect of neglecting Compton scattering (i.e.
for the case of absorption only) for the two lowest column densities.}
\end{figure}

\begin{figure}
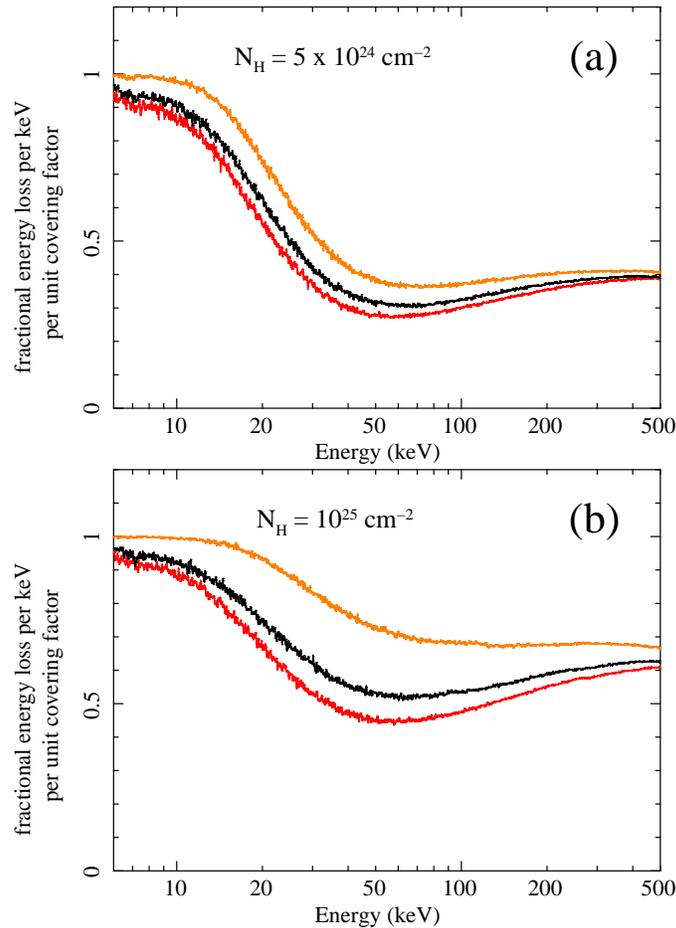

\centerline{
\psfig{figure=f2a.ps,width=9cm,angle=270}}
\centerline{
\psfig{figure=f2b.ps,width=9cm,angle=270}}
\caption[Zoom of Energy Loss Functions]{\footnotesize The dependence of the energy 
loss functions, \lossymp,
on the covering factor, $[\Delta\Omega/(4\pi)]=(a/c)$. Shown are curves for the two highest 
column densities in \figelossfuncp, $N_{H} = 5 \times 10^{24} \ \rm cm^{-2}$ 
and $10^{25} \ \rm cm^{-2}$. The dependence on $(a/c)$ for
$N_{H} = 10^{24} \ \rm cm^{-2}$ is negligible compared to
the statistical errors. Each curve has been normalized to the
particular value of $(a/c)$ for that curve in order to facilitate
comparison of the
{\it shape} of the functions. The values of the 
covering factor, $(a/c)$, in both (a) and (b)
are 0.1 (red), 0.5 (black), and 1.0 (brown). 
}
\end{figure}

The fractional energy loss curves in \figelossfunc and \figelossac
do not depend on the shape of the incident X-ray continuum 
spectrum because the curves give the {\it fractional} energy loss for 
monoenergetic photon injection. In order to calculate the absolute
energy losses the curves must be multiplied by the incident
continuum energy spectrum, as in equation~\ref{eq:abseloss}
(where $A$ is an arbitrary flux normalization).
Since the intrinsic X-ray continuum of AGNs can typically
be characterized by a power law with a photon index, $\Gamma$, in the
range 1.5--2.5, we can expect that the energy losses
due to Compton downscattering compared to absorption will be
diminished more for steeper spectra than for flatter spectra.

\begin{eqnarray}
\label{eq:abseloss}
L[(a/c),N_{H},E] & = &
D[(a/c),N_{H},E] A E^{-\Gamma+1} \ \ \ \ \ 
{\rm keV \ cm^{-2} \ s^{-1} \ keV^{-1}}
\end{eqnarray}

\section{Cumulative and total energy lost}
\label{cumloss}

So far we have examined the energy losses for monoenergetic injection.
We now examine the energy losses integrated over a range of energy.
We define the cumulative energy loss,
$CL[(a/c),N_{H},\Gamma,E]$ for a power-law
incident photon spectrum of the form $AE^{-\Gamma}$, between a lower energy,
$E_{L}$ and an energy $E$, as a fraction of the 
total incident energy in the energy range $E_{L}$ to some
upper energy, $E_{U}$. Thus, we have 

\begin{eqnarray}
\label{eq:ecumdef}
CL[(a/c),N_{H},\Gamma,E] & = & 
\frac{\int_{E_{L}}^{E}{L[(a/c),N_{H},E] \ dE}}
{\int_{E_{L}}^{E_{U}}{E^{-\Gamma+1} \ dE}}.
\end{eqnarray}

The absolute normalization of the incident continuum
of course cancels out in equation~\ref{eq:ecumdef}.
In the remainder of this paper we
use $E_{L}=0.5$~keV and $E_{U}=500$~keV.
The cumulative fractional energy loss function as defined in
equation~\ref{eq:ecumdef} is useful for showing the energy range
over which energy losses are most important (for a given set
of the parameters $(a/c)$, \nhp, and $\Gamma$). In \figelosscum
we show calculations of the function $CL[(a/c),N_{H},\Gamma,E]$ in
equation~\ref{eq:ecumdef} for
a covering factor of 0.5 and three different values of
$\Gamma$ (1.5, 1.9, and 2.5), each for two different values of \nh 
(red curves correspond
to $3 \times 10^{22} \rm \ cm^{-2}$ and
black curves correspond to $10^{25}  \rm \ cm^{-2}$).
These two extreme column densities correspond to 
equatorial Thomson depths of the torus of $\sim 0.024$ and
$\sim 8.1$ for the smaller and larger column respectively,
and are therefore representative of
a Compton-thin and a Compton-thick case respectively.
In \figelosscump, the total energy loss in the range 0.5--500~keV
as a fraction of the total incident energy in the same energy
band can be read off from the curves at $E=500$~keV. 

\begin{figure}
\centerline{
\psfig{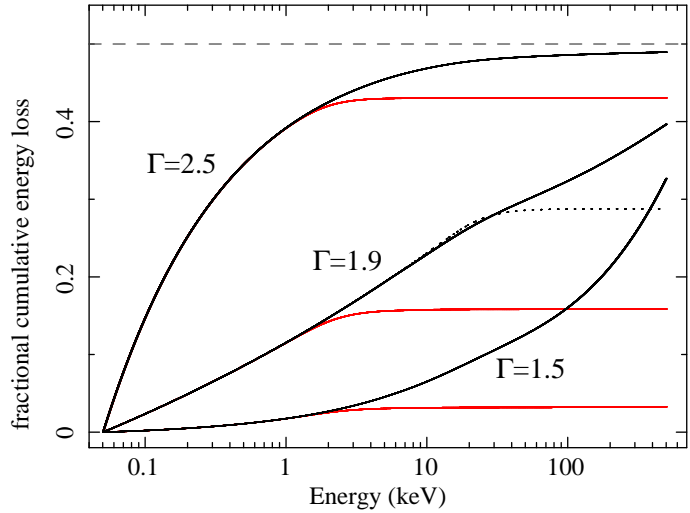}}
\caption[Cumulative Energy Loss Functions]{\footnotesize
The cumulative fractional energy loss in the torus as a function
of energy (i.e. from curves such as those in Fig.~1, integrated
from 0.05~keV up to an energy, $E$; see equation~\ref{eq:ecumdef}).
Pairs of curves are shown for a given incident power-law photon
index ($\Gamma$), for two column densities: $3 \times 10^{22} \ \rm cm^{-2}$
(red) and $10^{25} \ \rm cm^{-2}$ (black). Three pairs of curves
are shown, corresponding to $\Gamma = 1.5,1.9$, and 2.5. The
largest value that any of the curves can possibly have is equal to the
covering factor, $[\Delta\Omega/(4\pi)]=(a/c)$,
corresponding to {\it all} the energy
intercepted by the torus being captured. These
calculations were done for  $(a/c)=0.5$ and this value is shown
by the gray dashed line. The dotted curve shows the effect
of ignoring Compton scattering for the case of $\Gamma=1.9$ and
$N_{H} = 10^{25} \ \rm cm^{-2}$ (i.e. pure absorption). 
}
\end{figure}

Two results are immediately apparent from \figelosscump.
The first is that changing the column density by a factor of
more than 300, from $3 \times 10^{22} \ \rm cm^{-2}$
to $10^{25} \ \rm cm^{-2}$ produces a disproportionately small
change in the total integrated fractional energy loss, 
$CL[(a/c),N_{H},\Gamma,E_{U}]$. For $\Gamma=2.5$
the change in $CL$ is {\it only 14\% for more than two orders
of magnitude change in the column density}. For $\Gamma=1.9$ and
$1.5$ the change in CL between the two column densities is
only a factor of $\sim 2.5$ and $\sim 11$ respectively.
The latter still falls short by two orders of magnitude for the
change in $CL$ to be commensurate with the change in \nhp.
The second result that is apparent from \figelosscum is that 
for $\Gamma=2.5$ most of the incident energy has been deposited
in the torus below 10~keV, and the incident X-ray continuum
above 10~keV makes an insignificant contribution. On the
other hand, for $\Gamma=1.5$, most of the energy is deposited
{\it above} 10~keV. This is because
the shape of the incident spectrum
is critical on determining whether absorption or Compton scattering
dominates the energy losses. Steeper spectra have relatively
more photons at lower energies than flatter spectra so
absorption is correspondingly more important for steeper spectra.
In fact, \figelosscum shows that the total integrated energy loss
for $\Gamma=2.5$ is larger for the smaller column density than
it is for both $\Gamma=1.5$ and $\Gamma=1.9$ for the {\it larger} 
column density. In other words, the energy loss for the
$\Gamma=2.5$ {\it Compton-thin case is larger than the Compton-thick
case for} $\Gamma=1.5$ and $\Gamma=1.9$.
Finally, we also show in \figelosscum the result 
(dotted line) for $\Gamma=1.9$ when
Compton scattering is neglected (i.e. absorption only). 
It can be seen that for $\Gamma=1.9$, Compton scattering
increases the energy loss by $\sim 40\%$ compared to 
the case of absorption only.

\subsection{Dependence on column density}
\label{nhdepend}

In this section we examine the explicit dependence
of the total integrated fractional energy loss, \losstotfp, on \nhp.
\figtotvsnh shows this dependence
for three values of $\Gamma$ (1.5, 1.9, and 2.5)
and for each of these three values of $\Gamma$ two curves are shown,
corresponding to two values of the covering factor, $(a/c)=0.1$ 
(dotted) and $(a/c)=1.0$ (solid).
Note that the quantity that is actually plotted in \figtotvsnh
is \losstotfp/$(a/c)$, the total integrated fractional energy loss
per unit covering factor. The reason for this is to facilitate
a direct comparison of the {\it shapes} of the curves for 
different covering factors.
We see from \figtotvsnh that for $\Gamma=2.5$, the
shape of the total energy loss curves as a function of \nh
does not depend on the covering factor. For flatter spectra
there is some dependence that develops for column densities
greater than $\sim 10^{24} \rm \ cm^{-2}$ but even for
$\Gamma=1.5$ the difference between the $(a/c)=0.1$ and $(a/c)=1.0$
curves is no more than $\sim 25\%$ at the highest column density
($10^{25} \rm \ cm^{-2}$).

The most important result that \figtotvsnh shows is,
as already discussed above for \figelosscump, the very different behavior
of the energy loss functions as a function of \nh for different
values of $\Gamma$. For the steepest incident X-ray continuum
with $\Gamma=2.5$, the sensitivity of \losstotf to \nh is
less than $20\%$ for {\it a factor of $\sim 330$ change in} \nhp.
This is because for such a steep spectrum most of the energy
is in the soft X-ray band, and photons with such low energies
are readily absorbed by small column densities. Adding more 
column density then cannot make
much difference to the energy loss functions
if the bulk of the energy has already been deposited in the torus.
\figtotvsnh shows that even for $\Gamma=1.5$, the change in
\losstotf as \nh changes by a factor of $\sim 330$ is only just
over an order of magnitude.
Overall, \figtotvsnh shows that the total integrated fractional
energy loss is actually not very sensitive to \nh for any value
of $\Gamma$ or $(a/c)$.

\begin{figure}
\centerline{
\psfig{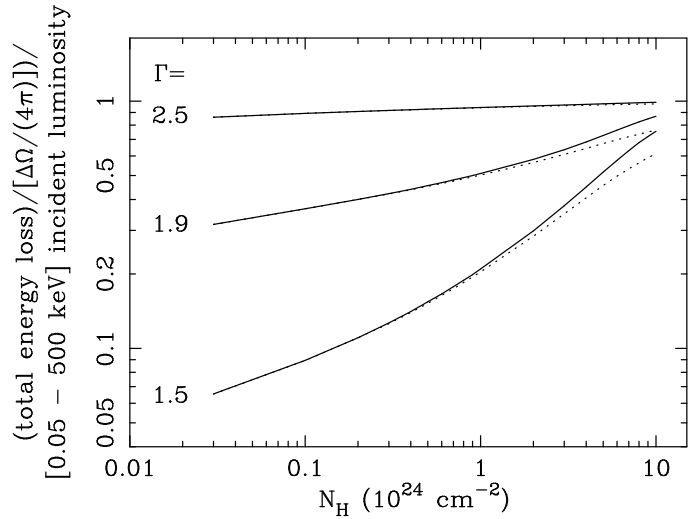}}
\caption[Column Density Dependence of Energy Loss]{\footnotesize
The total 0.05--500~keV integrated energy loss in the torus
as a fraction of the total integrated energy in the same bandpass,
as a function of the
torus equatorial column density, $N_{H}$.
Three pairs of curves
are shown, corresponding to photon indices
of the incident power-law spectrum, $\Gamma$, of 1.5, 1.9, and 2.5. 
Each pair corresponds to two different values of the covering
factor, $[\Delta \Omega/ (4\pi)]=(a/c)$, of 0.1 (dotted)
and 1.0 (solid). The curves have been divided by the covering
factor in order to facilitate a direct comparison of the
{\it shape} of each pair of curves. The functional
dependence on $N_{H}$ does not vary significantly with $(a/c)$.} 
\end{figure}

\subsection{Dependence on covering factor}
\label{covdepend}

In this section we examine the dependence of the total, integrated
fractional energy loss,
$CL[(a/c),N_{H},\Gamma,E_{U}]$ on the covering factor, $(a/c)$.
\figtotvsac shows curves of the total integrated fractional
energy loss versus $(a/c)$, for five different column densities
in the range $3 \times 10^{22} \ \rm cm^{-2}$ to
$10^{25} \ \rm cm^{-2}$, calculated for $\Gamma=1.9$.
It can be seen that the relationship between
$CL[(a/c),N_{H},\Gamma,E_{U}]$ and $(a/c)$ for a given 
value of \nh is very simple. In fact, as might be expected,
the relationship is linear for the
$N_{H}=10^{24} \ \rm cm^{-2}$ curve and for the curves
with smaller column densities, but there
is little departure from
linearity even for the $N_{H}=10^{25} \ \rm cm^{-2}$ curve.
The approximate linearity of $CL[(a/c),N_{H},\Gamma,E_{U}]$ as
a function of $(a/c)$ even in the Compton-thick regime is 
true for any $\Gamma$ in the range relevant to the present
study (range 1.5--2.5).
The significance of this result is that, since we have already 
seen that the dependence of $CL[(a/c),N_{H},\Gamma,E_{U}]$ on
\nh can be very weak (much weaker than linear), we can expect the
covering factor to play a role in the reprocessing of X-rays
to infrared emission that is at least, if not more important,
than that of the column density.

\begin{figure}
\centerline{
\psfig{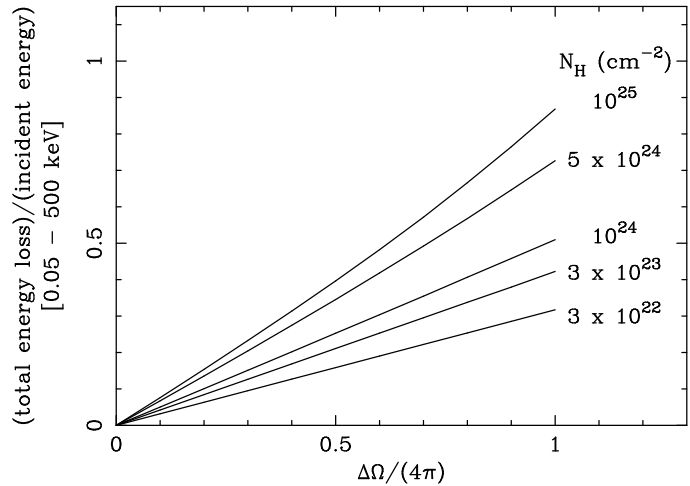}}
\caption[Covering Factor Dependence of Energy Loss]{\footnotesize
The total 0.05--500~keV integrated energy loss in the torus
as a fraction of the total integrated energy in the same bandpass,
as a function of the covering factor, 
$[\Delta \Omega/ (4\pi)]=(a/c)$. Curves are shown for five different
column densities, as indicated, for
an incident power-law photon index of $\Gamma=1.9$. It can be seen that the
total integrated fractional energy loss is approximately a linear
function of the covering factor.
}
\end{figure}

\subsection{Dependence on photon index}
\label{gamdepend}

In this section we examine the explicit dependence of the
total integrated fractional energy loss, \losstotfp, on 
the photon index of the incident power-law continuum, $\Gamma$.
\figtotvsgam shows this dependence 
for five different column densities
in the range $3 \times 10^{22} \ \rm cm^{-2}$ to
$10^{25} \ \rm cm^{-2}$, for $(a/c)=0.5$ (solid curves),
as $\Gamma$ varies from 1.4 to 2.6.
Also shown in \figtotvsgam are curves for $(a/c)=0.1$ and
$(a/c)=1.0$ for the two highest values of \nhp,
$5 \times 10^{24} \rm \ cm^{-2}$ (dotted) and $10^{25} \rm \ cm^{-2}$
(dashed). The energy loss has again been normalized by the
covering factor, $(a/c)$, so that the {\it shape} of the curves can
be directly compared for different values of the covering factor.
The curves for column densities lower than $5 \times 10^{24} \rm \ cm^{-2}$ 
in \figtotvsgam do not show a discernible difference for
different values of $(a/c)$.

We see from \figtotvsgam that the total fractional energy
loss functions for the {\it smallest} column densities
are actually the {\it most} sensitive to $\Gamma$.
For the lowest column density ($3 \times 10^{22} \ \rm cm^{-2}$),
\losstotf varies by a factor of $\sim 18$ as $\Gamma$ varies
between 1.4 and 2.6. However, for the highest column density
($10^{25} \ \rm cm^{-2}$), \losstotf varies by only 
a factor of $\sim 1.3$--1.7 (depending on the covering factor)
as $\Gamma$ varies between 1.4 and 2.6. The reason for the
larger sensitivity for smaller column densities is again that
the dominant energy loss mechanism is absorption of low-energy
photons in that regime. When the medium becomes Compton-thick,
energy losses due to Compton downscattering dominate and
multiple scatterings of high-energy photons tend to mitigate
the sensitivity to $\Gamma$. 

\begin{figure}
\centerline{
\psfig{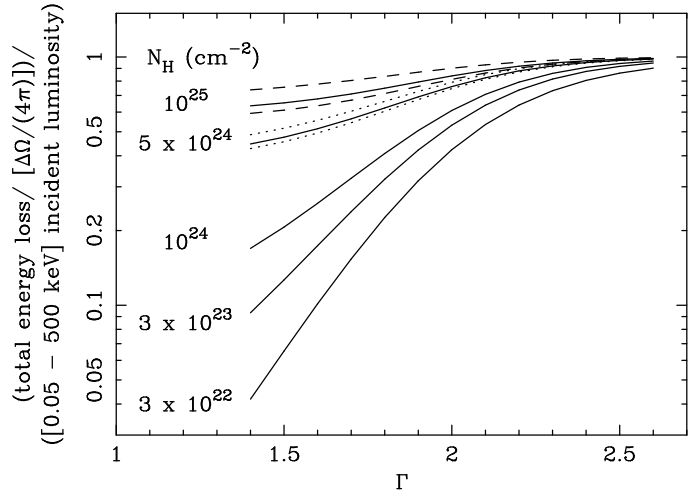}}
\caption[Gamma Dependence of Energy Loss]{\footnotesize
The total 0.05--500~keV integrated energy loss in the torus
as a fraction of the total integrated energy in the same bandpass,
as a function of the
photon index of the incident power-law spectrum ($\Gamma$),
for five different values of the torus equatorial column density,
as indicated. The solid lines correspond to calculations with
a covering factor, $[\Delta \Omega/ (4\pi)]=(a/c)$ of 0.5.
The dotted lines correspond, for $N_{H} = 5 \times 10^{24} \ \rm cm^{-2}$,
to $(a/c)=0.1$ (lower curve), and $(a/c)=1.0$ (upper curve).
The dashed lines correspond, for $N_{H} = 10^{25} \ \rm cm^{-2}$,
to $(a/c)=0.1$ (lower curve), and $(a/c)=1.0$ (upper curve).
All of the curves have been divided by the covering 
factor in order to facilitate a direct comparison of their
{\it shape}.
}
\end{figure}

\section{Energy loss as a fraction of the observed to intrinsic X-ray luminosity ratio}
\label{irxproxy}

So far we have seen that of the three critical parameters of
the system consisting of the
X-ray reprocessor and its illuminating continuum
(covering factor, \nhp, and $\Gamma$), the 
energy deposited in the medium that is potentially available
for reprocessing into infrared emission, 
is {\it least} sensitive to \nhp. This already does not look
promising for supporting the idea that the column density of
the obscuring matter in AGNs could be constrained by the
infrared emission. However, it is not the absolute 
infrared luminosity that is claimed in the literature to
be the indicator of column density, but the ratio of the infrared
luminosity to the {\it observed} X-ray luminosity. Since the
latter has a strong dependence on the line-of-sight column density,
it might be supposed that the ratio does in fact provide a
good indicator of the column density. In order to address this
question, we constructed a proxy for the infrared to X-ray luminosity
ratio that is due to X-ray energy from the illuminating
continuum deposited in the obscuring matter. Specifically,
using our \mcarlo results,
we calculated the ratio of the total integrated fractional
energy loss, \losstotf (see equation~\ref{eq:ecumdef}) to the
ratio of the observed to intrinsic X-ray luminosity in the 2--10~keV
band. We call this quantity \proxyratiop, which can be written as

\begin{eqnarray}
\label{eq:proxydef}
PX[(a/c),N_{H},\Gamma] & = &
\frac{CL[(a/c),N_{H},\Gamma,E_{U}]}{[L_{\rm observed}/L_{\rm intrinsic}]_{\rm 2-10 \ keV}}.
\end{eqnarray}

The quantity \losstotf represents only the energy {\it potentially} available
for reprocessing into infrared emission so the quantity \proxyratio is
really an upper limit for a given set of parameters.
However, this does not impact the basic test that we want to
perform. We need to establish whether the ratio \proxyratio is
sensitive enough to \nh that it could distinguish between
a Compton-thin and a Compton-thick AGN, despite the sensitivity
of that ratio to $\Gamma$ and to the covering factor. Deep X-ray
survey data on AGNs do not have a sufficiently high
signal-to-noise ratio to unambiguously determine $\Gamma$ and $(a/c)$.
Both parameters are highly model-dependent and
degenerate with each other and with
other model parameters for the signal-to-noise ratio that
is typical of these deep survey AGN data
(e.g., see discussion in Georgantopoulos \etal 2009). In the
least constrained scenarios, X-ray spectroscopy is not even
possible and inferences are made from X-ray hardness ratios
alone, which carry an even greater degree of degeneracy (e.g.,
Polletta \etal 2006). We also note that covering factors
deduced independently from other wavebands (e.g., the
dust covering factor) are not necessarily equal to the
X-ray covering factor.
Therefore, in the absence of knowledge of $\Gamma$ and
the covering factor, 
if the theoretical ratio
\proxyratio {\it is} sufficiently sensitive to distinguish between a
Compton-thin and a Compton-thick AGN, then the observed infrared to
X-ray luminosity ratio {\it might} be a useful indicator of the
column density. On the other hand, if \proxyratio {\it is not}
sufficiently
sensitive to distinguish between a
Compton-thin and a Compton-thick AGN, then the observed infrared to
X-ray luminosity ratio {\it is not} an indicator of the column density. 

\begin{figure}
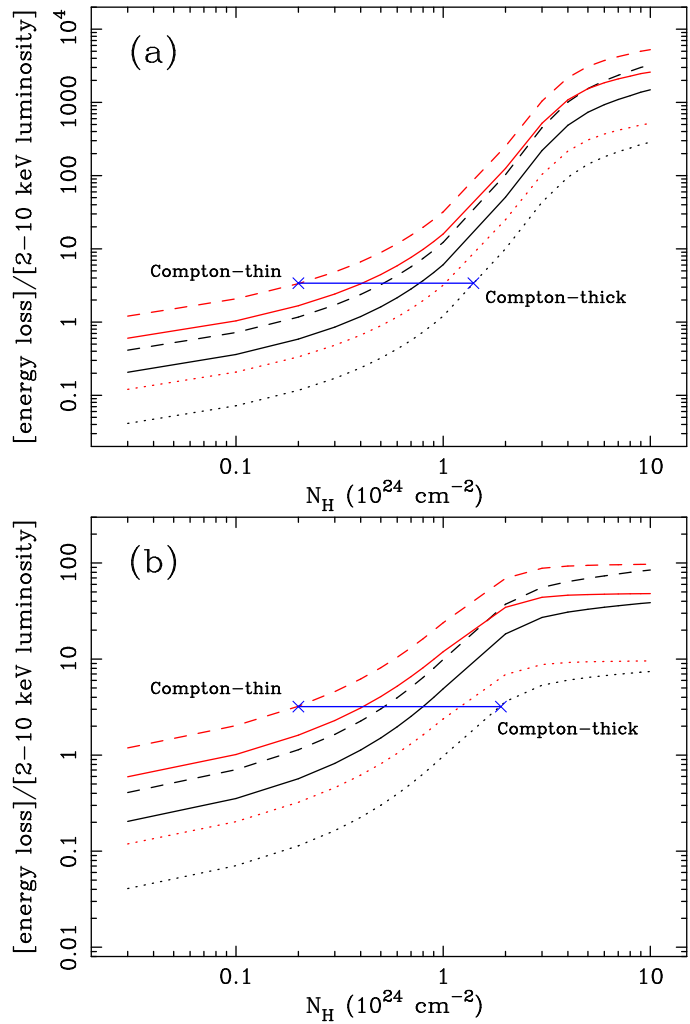

\centerline{
\psfig{figure=f7a.ps,width=9cm,angle=270}}
\centerline{
\psfig{figure=f7b.ps,width=9cm,angle=270}}
\caption[Proxy for IR/X Ratio]{\footnotesize
(a)
The ratio of the total 0.05--500~keV integrated fractional energy loss 
as fraction of the 2--10~keV observed to
intrinsic luminosity ratio, for
a torus viewed edge-on, as a function of the column density.
This ratio, \proxyratio (see equation~\ref{eq:proxydef}),
is a proxy for the ratio of 
the fraction of the total X-ray luminosity that could
be reprocessed into the infrared band,
to the 2--10~keV X-ray observed to intrinsic luminosity ratio. 
Curves are shown for two
values of the photon index of the incident power-law spectrum,
$\Gamma = 1.9$ (black) and 2.5 (red). For each value
of $\Gamma$, the curves are calculated
for three values of the covering factor, $(a/c)=\Delta \Omega/ (4\pi)=0.1$
(dotted), 0.5 (solid), and 1.0 (dashed).
The blue crosses joined by the blue horizontal line show that
the proxy \proxyratio is not a good indicator of $N_{H}$ because,
in this example, a Compton-thin source with $N_{H}=2 \times 10^{23}
\ \rm cm^{-2}$ and a Compton-thick source with $N_{H}=1.4 \times 
10^{24} \ \rm cm^{-2}$ have the {\it same} value of the proxy.
This is because the dependence of the proxy on $\Gamma$ and
the covering factor can be greater than the dependence on $N_{H}$. 
(b) As (a) but this time the 2--10~keV luminosity included
a power-law continuum due to scattering in an optically-thin
zone surrounding the torus, corresponding to 1\% of the direct
(incident) X-ray continuum. The dependence of the
proxy \proxyratio on $N_{H}$ is now even weaker.
The blue crosses joined by the blue horizontal line show that
a Compton-thin source with $N_{H}=2 \times 10^{23} 
\ \rm cm^{-2}$ and a Compton-thick source with $N_{H}=1.9 \times  
10^{24} \ \rm cm^{-2}$ have the {\it same} value of the proxy. 
}
\end{figure}

Any additional infrared emission due to dust emission can
only serve to {\it reduce} the usefulness of the
observed infrared to
X-ray luminosity ratio as an indicator of column density.
There is at least one more complication. That is, we know from
studies of bright AGNs that even in obscured sources the
X-ray spectrum often shows an {\it unobscured} continuum component
that is due to optically-thin electron scattering in an 
extended zone surrounding the central engine. Typically, this
component can be characterized by a power-law continuum that has
a luminosity that is $\sim 0.02$--5\% of
the direct (incident) continuum (e.g., Turner \etal 1997). 
However, in X-ray deep survey data the
signal-to-noise ratio of the spectra for individual AGN is
too poor to constrain this component. The resulting spectrum
can then appear to have a smaller column density than 
that which is actually
obscuring the primary continuum if the optically-thin scattered
continuum is not included in the model (and if it is, some
assumptions have to usually be made about the pertinent parameters).
In calculating the observed to intrinsic X-ray luminosity ratio
in equation~\ref{eq:proxydef} we can include the
optically-thin scattered continuum component, with its photon index
and luminosity relative to the direct (incident) continuum as additional
parameters. We note that the observed to intrinsic X-ray luminosity
ratio depends on the orientation at which the torus is observed.
In the following examples we will use an edge-on orientation
because the the observed to intrinsic X-ray luminosity
ratio is most sensitive to \nh when the torus is observed edge-on.
Since we have already found that the energy losses are not
very sensitive to \nhp, we are looking for ways that \proxyratio
could recover some sensitivity to \nh and the edge-on orientation
is the most appropriate for this purpose.  

In \figproxyvsnh we show calculations of \proxyratio from
our \mcarlo results. The curves in \figproxyvsnha were calculated
with no optically-thin scattered continuum component and the
curves in \figproxyvsnhb were calculated with the inclusion
of a continuum component with the same value of $\Gamma$ as the
primary continuum but with only 1\% of the luminosity of the
direct (incident) continuum. The curves shown in \figproxyvsnh
were calculated for three values of the covering factor,
$(a/c)=0.1$ (dotted lines), $(a/c)=0.5$ (solid lines), and $(a/c)=1.0$
(dashed lines), and two values of $\Gamma$ (1.9 and 2.5,
corresponding to the black and red curves respectively).

We see from \figproxyvsnh that the sensitivity of \proxyratio
to the covering factor and $\Gamma$ is still very strong and
can in fact override the weak dependence of  \proxyratio on \nhp.
We also see that the effect of including the optically-thin
scattered continuum, even at the level of 1\%, significantly
weakens the sensitivity of  \proxyratio to \nhp.
For the highest column densities, in the Compton-thick regime,
we see that \proxyratio drops by two orders of magnitude
due to the optically-thin scattered continuum component,
for the same values of $\Gamma$, $(a/c)$, and \nhp.

In both \figproxyvsnha and \figproxyvsnhb we illustrate
one example in each case of a situation in which 
a Compton-thin AGN can give the same value of the ratio
\proxyratio as a Compton-thick AGN. 
The blue crosses joined by the blue horizontal lines show that the
infrared to X-ray luminosity 
proxy is {\it not} a good indicator of $N_{H}$.
In \figproxyvsnha
the example shows that a Compton-thin source with $N_{H}=2 \times 10^{23}
\ \rm cm^{-2}$ and a Compton-thick source with $N_{H}=1.4 \times 
10^{24} \ \rm cm^{-2}$ have the {\it same} value of the proxy.
In other words, a Compton-thin AGN with a steep spectrum and a 
high covering factor could give the same value of \proxyratio
as a Compton-thick AGN with a flatter spectrum and a smaller covering
factor. In \figproxyvsnhb
the blue crosses joined by the blue horizontal line show that
a Compton-thin source with $N_{H}=2 \times 10^{23} 
\ \rm cm^{-2}$ and a Compton-thick source with $N_{H}=1.9 \times  
10^{24} \ \rm cm^{-2}$ have the {\it same} value of the proxy. 

A general conclusion that can be drawn from \figproxyvsnh is
that if we have a large sample of AGNs from a deep X-ray survey and
examine the distribution of the ratio of observed infrared 
luminosity to X-ray luminosity, we could say that the
AGNs with the highest value of that ratio {\it might} be
Compton-thick candidates and/or heavily contaminated by
dust emission. For the bulk of the AGNs in the distribution,
the infrared luminosity to X-ray luminosity of a given
AGN would tell us
very little about the column density in the absence of
information on the intrinsic X-ray continuum, the optically-thin scattered
continuum, and the covering factor. 
This realization
has already become apparent to Georgantopoulos \etal (2010)
who have observationally identified some 
X-ray selected ``infrared excess''
AGNs that they call DOGs (Dust Obscured Galaxies) to
be {\it Compton-thin}.
Our theoretical results explain the reasons why the
infrared luminosity to X-ray luminosity in AGNs is not a good
indicator of the column density.

Our calculations have assumed a uniform distribution of matter,
but the toroidal matter 
distribution may be clumpy (e.g., see Elitzur 2008, and references therein).
In the optically-thin limit, a clumpy matter distribution will give
the same results for the same {\it actual} covering factor, except that
the covering factor is no longer an indicator of the solid
angle subtended by the reprocessor at the X-ray source. An additional
parameter, a filling factor, would be required to derive a
relation between the solid angle and the covering factor.
In the limit that each clump of matter is Compton thick, the total
energy loss to the torus would be {\it less} than the corresponding 
uniform matter distribution with the same covering factor and mean
column density (for the same incident X-ray continuum). This is
because more lines-of-sight are available for X-ray photons to
escape from the surface of a Compton thick clump after one or more
scatterings (which may occur in more than one clump). Therefore
the overall effect of clumpiness is to make the energy loss
even less sensitive to the mean column density of the matter distribution
because in the optically-thin limit the energy losses are
the same but in the Compton-thick limit the losses are less for
a clumpy distribution compared to a uniform matter distribution.
Thus, the principal conclusions of the present paper
still hold for a clumpy matter distribution.

\section{Summary}
\label{summary}

The aim of this paper was to investigate the theoretical
foundation of the idea that the observed 
mid-infrared luminosity to X-ray luminosity ratio in AGNs that are
found in deep X-ray surveys can be used as a proxy for
the column density of obscuring matter, which therefore might be
useful for identifying
Compton-thick candidates. The basis for this is
that energy lost by the X-ray continuum to the
obscuring matter appears as reprocessed infrared emission.  
In order for the proxy to be viable a {\it minimal} requirement
is that the energy losses due to absorption and Compton scattering
in the obscuring matter should be sensitive to the
column density and {\it insensitive to other key physical parameters
of the system}. 
By means of \mcarlo simulations we found that the energy
deposited in the obscuring matter and available for
reprocessing into infrared emission has a sensitivity to
the shape (steepness) of the incident X-ray continuum  
and the covering factor of the intercepted material that
is far greater than the sensitivity to the column density.
As a result, we found that the observed infrared to X-ray
luminosity ratio for a Compton-thin AGN could be just as large
as that for a Compton-thick AGN. The signal-to-noise
ratio of the X-ray spectra of AGNs currently
found in deep X-ray surveys
is not sufficiently high to constrain the intrinsic X-ray
spectrum and the covering factor of the reprocessor 
well enough to enable the infrared to X-ray
luminosity ratio to be used as a reliable indicator of column density.
If the infrared emission has a significant contribution
from dust emission, for example due to starburst activity, 
the reliability of that indicator
can only be worse. 
Our conclusion regarding the difficulty in the use of
the mid-infrared to X-ray luminosity ratio for distinguishing
between Compton-thin and Compton-thick AGNs
 is independent of the
geometry of the obscuring matter in AGNs. It is also independent
of the angular distribution of the intrinsic X-ray continuum
emission. This is because the dominant energy loss mechanism changes
from absorption to Compton scattering if the
slope of the X-ray continuum
is changed from having a power-law photon index of 2.5 to 1.5
(the typical range in AGNs), and this fact is independent of geometry.
Further, the energy deposited in the obscuring matter trivially
depends on the covering factor, and this fact is also independent
of geometry.

Acknowledgments \\
Partial support (TY) for this work was provided by NASA through \chandra Award
AR8-9012X, issued by the Chandra X-ray Observatory Center,
which is operated by the Smithsonian Astrophysical Observatory for and
on behalf of the NASA under contract NAS8-39073.
Partial support from NASA grants NNX09AD01G and NNX10AE83G is also
(TY) acknowledged.

\end{document}